\documentclass[aps,prl,reprint, amsmath, amssymb,superscriptaddress]{revtex4-1}

\usepackage{bm}
\usepackage[retainorgcmds]{IEEEtrantools}
\usepackage{graphicx}
\usepackage{mathrsfs}
\usepackage{amsmath}
\usepackage{amssymb}
\usepackage{color}
\usepackage{amsfonts,txfonts}
\usepackage{nicefrac}

\newcommand{\ud}{\,\mathrm{d}}

\DeclareMathOperator{\tr}{tr}

\newcommand{\nn}{\bar{n} + {1}/{2}}
\newcommand{\Tr}{\rm Tr}

\begin{document}

\title{The Wigner entropy production rate}
\date{\today}
\author{Jader P. Santos}
\affiliation{Universidade Federal do ABC,  09210-580 Santo Andr\'e, Brazil}
\author{Gabriel T. Landi}
%\email{gtlandi@if.usp.br}
\affiliation{Instituto de F\'isica da Universidade de S\~ao Paulo,  05314-970 S\~ao Paulo, Brazil}
\author{Mauro Paternostro}
\affiliation{Centre for Theoretical Atomic, Molecular and Optical Physics,
School of Mathematics and Physics, Queen's University Belfast, Belfast BT7 1NN, United Kingdom}

\begin{abstract}
The characterization of irreversibility in general quantum processes is an open problem of increasing technological relevance. 
Yet, the tools currently available to this aim are mostly limited to the assessment of dynamics induced by equilibrium environments, 
a situation that often does not match the reality of experiments at the microscopic and mesoscopic scale. We propose a theory of 
irreversible entropy production that is suited for quantum systems exposed to general, non-equilibrium reservoirs. We illustrate our 
framework by addressing a set of physically relevant situations that clarify both the features and the potential of our proposal. 
\end{abstract}
\maketitle{}

%%%%%%%%%%%%%%%%%%%%%%%%%%%%%%%%%%%%%%
%
%
%		INTRODUCTION
%
%
%%%%%%%%%%%%%%%%%%%%%%%%%%%%%%%%%%%%%%

\emph{Introduction - }The entropy of an open system, unlike the energy, does not satisfy a continuity equation: in addition to  entropic fluxes exchanged with the environment, some entropy may also be produced within the system. 
This contribution is called the \emph{entropy production} and, according to the second law of thermodynamics, it is always non-negative, being zero only when the system and the environment are in thermal equilibrium. 
It therefore serves as a measure of the irreversibility of a physical process and may be used to characterize  non-equilibrium systems in a broad range of situations  and across all length scales. 
In symbols, if $S$ is the entropy of the system, then its rate of change may always be written as 
\begin{equation}\label{single:S_sep}
\frac{\ud S}{\ud t} = \Pi(t)- \Phi(t)
\end{equation}
where $\Pi \geq 0$ is the \emph{entropy production rate} and $\Phi$ is the \emph{entropy flux rate}, from the system to the environment. 
%The entropy production rate $\Pi$ is expected to be non-zero as long as the system is out of equilibrium. 
%This includes transient states and  non-equilibrium steady-states (NESSs), where $\ud S/\ud t= 0$ and therefore $\Pi = \Phi > 0$. 
The quantities $\Pi$ and $\Phi$ are not direct observables and must therefore be related to experimentally accessible  quantities  via a theoretical framework. 
Unfortunately, a unified approach for this is still lacking.

In the past decades, several theories of entropy production have been developed in different contexts.
The most prominent example is  Onsager's  theory of chemical kinetics \cite{Onsager1931,*Onsager1931a,Machlup1953,DeGroot1961,Tisza1957}, where the entropy production rate is  related to particle and energy currents.
Another widely used framework is that of Schnakenberg \cite{Germany1976,Tome2012}, which relates the entropy production rate with the transition rates of a system  governed by a master equation. 
The generalization to other classical stochastic processes, such as dynamics described by a Fokker-Planck equation, have also been addressed \cite{Tome2010,Spinney2012,Landi2013b}.

The extension of these results to mesoscopic systems came into relevance  with the discovery by  Gallavotti, Cohen and collaborators \cite{Evans1993,*Evans1994,Gallavotti1995b,*Gallavotti1995} that  the total entropy production $\Sigma$ of a process, when interpreted as a fluctuating quantity of the system's stochastic trajectory, satisfies a fluctuation theorem of the form $\langle e^{-\Sigma}\rangle = 1$, which is valid for processes arbitrarily far from equilibrium. 
Similar results were found  by Jarzynski \cite{Jarzynski1997,*Jarzynski1997a} and Crooks \cite{Crooks1998,*Crooks2000}  for systems undergoing a work protocol, where the entropy production is  proportional to the irreversible work. 
These developments and, in particular, their extensions to quantum systems, have shown that in meso and microscopic systems, quantum fluctuations may play a prominent role in non-equilibrium processes. 
 
Quantum systems also open up the possibility for exploring more general reservoirs, such as dephasing and squeezed baths~\cite{Clark2016}. The description of these systems extends beyond the usual paradigms of equilibrium environments. 
Despite the lack of equilibrium at the bath level, one should still be able to characterize processes by their irreversibility and entropy production. There is thus a strong need for the identification of suitable tools that are able to characterize non-equilibrium processes in a broad class of settings. 

The goal of this paper is to derive a theory of entropy production  that  is applicable to quantum systems subject to more general reservoirs.
Differently from existing theories, instead of using the von Neumann entropy $S_\text{vN} = - \tr(\rho\ln\rho)$, we shall  characterize the irreversibility using the R\'enyi-2 entropy $S_2 = - \ln \tr\rho^2$, where $\rho$ is the density matrix of the system. 
Both entropies have a similar behavior when used to characterize  disorder.
However, the R\'enyi-2 entropy is much more convenient to manipulate, since it is simply related to the purity  $\tr\rho^2$ of the state.
Recently, there has been several papers linking the general R\'enyi-$\alpha$ entropies $S_\alpha=(1-\alpha)^{-1}\ln\Tr(\rho^\alpha)$  to the thermodynamic properties of quantum systems, from the formulation of general fluctuation theorems to the derivation of a family of second laws of thermodynamics~\cite{Wei,Misra2015,Brandao2015}. 
Remarkably, all R\'enyi-$\alpha$ entropies tend asymptotically to the von Neumann one in the classical limit, corroborating their use in reformulating the theory of thermodynamic irreversibility.
%It is in this sense that our investigation on a formulation of entropy production in terms of $S_2$ should be assessed here. 
The subtleties implied by the differences between the von Neumann and R\'enyi entropies has been stressed in Ref.~\cite{Abe2016}.

In this paper, we shall focus on bosonic systems characterized by Gaussian states. 
In this case, the expression for $S_2$ is written simply as 
$S_2 = \frac{1}{2} \ln |\Theta|$, where $\Theta$ is the covariance matrix  \cite{Ferraro2005}.
Moreover, as shown in Ref.~\cite{Adesso2012}, $S_2$ coincides up to a constant with the  \emph{Wigner entropy}~
\begin{equation}\label{single:S}
S = - \int \ud^2\alpha\; W(\alpha^*,\alpha) \ln W(\alpha^*,\alpha).
\end{equation}
where $W(\alpha^*,\alpha)$ is the Wigner function and the integral is over the complex plane (as the state is Gaussian, $W>0$ and hence $S$ is real).
This link between $S$ and $S_2$ allows for a fundamental simplification of the problem of characterizing entropy production, as one can map the  open system dynamics into  a Fokker-Planck equation for $W$ and hence employ tools of classical stochastic processes 
to obtain  simple expressions for  $\Pi$ and $\Phi$.
This idea was already used in 
Refs.~\cite{Brunelli2016a,Brunelli2016} via a quantum-to-classical correspondence to treat the case of simple heat baths.
Here, instead, we present a full quantum mechanical treatment and show how to extend the framework to treat  squeezed and dephasing reservoirs.
The generalization to other types of baths is straightforward. 

We shall assume that the system is modeled by a Lindblad master equation of the form 
\begin{equation}\label{single:master}
\partial_t\rho= - i [H, \rho] + \mathcal{D}(\rho),
\end{equation}
where $\rho$ is the density matrix of the system, $H$ is its Hamiltonian and ${\cal D}(\rho)$ describes the process arising from its coupling to the external reservoir. 
Let $\rho^*$ denote the target state of $\mathcal{D}(\rho)$ (for thermal baths $\rho^* = \rho_{\text{eq}} = e^{-\beta H}/Z$). 
In Refs.~\cite{Spohn1978,Breuer2003,Breuer2007,Deffner2011}, it was shown that the von Neumann entropy  production rate can be defined as 
\begin{equation}\label{Pi:Breuer}
\Pi_{\text{vN}}= - \partial_t K_{\text{vN}}(\rho |  \rho^*),
\end{equation}
where $K_{\text{vN}}(\rho | \rho^*) = \tr[\rho \ln (\rho/\rho^*)]$ is the von Neumann relative entropy.
Eq.~(\ref{Pi:Breuer}) satisfies several properties expected from an entropy production.
First, $\Pi_{\text{vN}}\ge0$,  with the equality holding only for $\rho = \rho^*$. 
Second, for thermal baths, the corresponding total entropy production, when interpreted as a stochastic quantity,   satisfies an integral fluctuation theorem \cite{Deffner2011}. 
Finally, Eq.~(\ref{Pi:Breuer}) may be factored in the form of Eq.~(\ref{single:S_sep}), with $S=S_\text{vN}$ and
\begin{equation}\label{Phi_classical}
\Phi_{\text{vN}}(t) = - \frac{1}{T} \tr\bigg[ H \mathcal{D}(\rho)\bigg] := \frac{\Phi_E}{T},
\end{equation}
where $\Phi_E$ is the energy flux from the system to the environment. 
This is a well known  result of classical thermodynamics, relating heat and entropy flux.

Despite their clear physical interpretation,  Eqs.~(\ref{Pi:Breuer}) and (\ref{Phi_classical}) suffer from the problem that they diverge in the limit $T\to 0$. This is related to the divergence of the relative entropy when the reference state tends to a pure state \cite{Abe2003,Audenaert2013}. 
This divergence is clearly an inconsistency of the theory.  
The limit $T\to 0$ is frequently used in quantum optics and  the dynamics is known to be  well behaved and to correctly reproduce experimental results in several situations. 
In fact, even $\ud S/\ud t$ remains finite in this limit, and only $\Pi$ and $\Phi$ diverge. 
In the past, several attempts have been made to overcome this problem~\cite{Frank2013,Muller-Lennert2013,Audenaert2013,Abe2003,Esposito2010a,Pucci2013} but a consistent theory is still lacking.
To obtain a framework which does not suffer from this deficiency is another motivation for this paper.
As we will show, using the R\'enyi-2/Wigner entropy avoids this problem entirely.

%%%%%%%%%%%%%%%%%%%%%%%%%%%%%%%%%%%%%%
%
%
%		SINGLE HARMONIC OSCILLATOR	
%
%
%%%%%%%%%%%%%%%%%%%%%%%%%%%%%%%%%%%%%%

\emph{Thermal bath - }We begin the construction of our formalism by considering a single bosonic mode with  $H = \omega (a^\dagger a+{1}/{2})$ and dissipator
\begin{equation}
\label{single:D}
\mathcal{D}(\rho) = \gamma(\bar{n}+1)\left[ a \rho a^\dagger - \frac{1}{2} \{a^\dagger a, \rho\}\right] + \gamma \bar{n} \left[ a^\dagger \rho a - \frac{1}{2} \{ a a^\dagger, \rho\}\right].
\end{equation}
Here $\gamma$ is the damping rate of the oscillator and $\bar{n} = (e^{\beta \omega}-1)^{-1}$ is the mean number of excitations in the bath ($\beta = 1/T$ is its inverse temperature). 
The target state of this dissipator is the Gibbs thermal state $\rho^* = \rho_{\text{eq}} = (1-e^{-\beta \omega})e^{-\beta \omega a^\dagger a}$.

We define the Wigner function of the system as 
\begin{equation}\label{single:Wdef}
W(\alpha^*,\alpha) = \frac{1}{\pi^2} \int\ud^2 \lambda \; e^{- \lambda \alpha^* + \lambda^* \alpha} \tr\left\{ \rho \; e^{\lambda a^\dagger - \lambda^* a}\right\},
\end{equation}
where $\lambda$ and $\alpha$ are phase space variables. %the integral is over the entire complex plane. 
Using standard operator correspondences, Eq.~(\ref{single:master}) 
can be translated into the Fokker-Planck equation
\begin{equation}\label{single:FP}
{\partial_t W} = - i \omega\bigg[ \partial_{\alpha^*} (\alpha^* W) - \partial_\alpha (\alpha W)\bigg] + \mathcal{D}(W),
\end{equation}
where the dissipative part is written as a divergence in the complex plane:
\begin{equation}\label{single:DW_J}
\mathcal{D}(W) = \partial_\alpha J(W) + \partial_{\alpha^*} J^*(W),
\end{equation}
with
\begin{equation}\label{single:J}
J(W) = \frac{\gamma}{2} \bigg[ \alpha W + (\overline{n}+1/2) \partial_{\alpha^*} W\bigg].
\end{equation}
Eq.~(\ref{single:FP}) is a continuity equation in the complex plane. 
Hence, $J(W)$ can be interpreted as the \emph{irreversible component of the  probability current}. 
This picture is further corroborated by the fact that $J(W)$ will be zero \emph{only} in the thermal state $W_{\text{eq}}=\frac{1}{\pi(\nn)}\exp[{ - \frac{|\alpha|^2}{\nn}}]$; i.e., $J(W_{\text{eq}}) = 0$.  
This statement is stronger than $\mathcal{D}(W_\text{eq}) = 0$ as it implies  that the thermal equilibrium state is not only a fixed point of the dissipative dynamics, but also the state where all  probability currents vanish identically.

 Having defined the Wigner entropy as in Eq.~(\ref{single:S}), we now define the Wigner entropy production rate as 
\begin{equation}\label{single:Pi:relative}
\Pi = - \partial_t K(W(t) || W_\text{eq}),
\end{equation}
where
%\begin{equation}\label{single:relative}
$K(W||W_\text{eq}) = \int\ud^2\alpha \; W \ln W/W_\text{eq}$
%\end{equation}
is the Wigner relative entropy. For a bipartite Gaussian state, this coincides (up to a constant) with the R\'enyi-2 mutual information \cite{Adesso2012}.
Inserting the  Fokker-Planck Eq.~(\ref{single:FP}) in Eq.~\eqref{single:Pi:relative} and  integrating by parts we get
\begin{equation}\label{single:Pi_2terms}
\Pi = - \int\ud^2\alpha\; \mathcal{D}(W) \ln (W /W_\text{eq}).
\end{equation}
Next we use Eq.~(\ref{single:DW_J}) and integrate by parts again to obtain
\begin{equation}
\Pi = \int \ud^2\alpha \bigg\{ J \bigg(\frac{\partial_\alpha W}{W}- \frac{\partial_\alpha W_\text{eq}}{W_\text{eq}}\bigg) + \alpha\to\alpha^*
%J_{\alpha^*} \bigg(\frac{\partial_{\alpha^*} W}{W}- \frac{\partial_{\alpha^*} W_\text{eq}}{W_\text{eq}}\bigg)
\bigg\}.
\end{equation}
Finally one notes that, from Eq.~(\ref{single:J})
\begin{equation}
\frac{\partial_\alpha W}{W}- \frac{\partial_\alpha W_\text{eq}}{W_\text{eq}} = \frac{2J^*}{\gamma(\nn)} \frac{1}{W}.
\end{equation}
Therefore, we conclude that the entropy production rate may be written as 
\begin{equation}\label{single:Pi}
\Pi= \frac{4}{\gamma(\nn)}  \int\ud^2 \alpha \frac{|J(W)|^2}{W}.
\end{equation}
This quantity is always non-negative (as we take $W>0$) and null only at thermal equilibrium, which are precisely the properties expected from an entropy production rate. 

Going back to Eq.~(\ref{single:Pi_2terms}), the term proportional to $\mathcal{D}(W) \ln W$ is precisely $\ud S/\ud t$, with $S$ defined in Eq.~(\ref{single:S}). 
Hence, comparing with Eq.~(\ref{single:S_sep}) we find that the remainder must be the entropy flux rate. 
\[
\Phi = \int \ud^2\alpha \; \mathcal{D}(W) \ln W_\text{eq} = \frac{\gamma}{\nn} \int\ud^2\alpha \; |\alpha|^2 W - \gamma,
\]
where, in the last line, we integrated by parts and substituted the formulas for $\mathcal{D}(W)$ and $W_{\text{eq}}$. 
Since $\int\ud^2 \alpha \; |\alpha|^2 W = \langle a^\dagger a \rangle + {1}/{2}$ we finally conclude that 
\begin{equation}\label{single:Phi_3}
\Phi = \frac{\gamma}{\nn} (\langle a^\dagger a \rangle - \bar{n}).
\end{equation}
Eqs.~(\ref{single:Pi}) and (\ref{single:Phi_3}) are the main results for the Wigner entropy production and entropy flux rate. 
Eq.~(\ref{single:Phi_3}) in particular is very useful, as it relates the entropy flux rate to a simple expectation value. 

%We can also obtain a simpler formula for $\Pi$ by using the specific structure of Gaussian states. 
%Let $\bm{u} = (a,a^\dagger)$. We define the covariance matrix as $\Theta_{i,j} = \frac{1}{2} \langle \{u_i, u_j^\dagger \} \rangle - \langle u_i \rangle \langle u_j^\dagger \rangle$. 
%A straightforward calculation then shows that 
%\begin{equation}\label{single:Pi_2}
%\Pi = \Phi - \gamma + \gamma(\nn) \frac{\Theta_{11}}{|\Theta|}.
%\end{equation}
%which gives $\Pi$ in terms of $\Phi$ and elements of $\Theta$.

On the other hand, the energy flux rate may be computed from Eq.~(\ref{single:master}) and reads $\Phi_E = \gamma \omega (\langle a^\dagger a \rangle - \bar{n})$.
We thus conclude that the entropy flux rate and the energy flux rate are related by 
\begin{equation}
\Phi = \frac{\Phi_E}{\omega (\nn)}.
\end{equation}
When $T \gg \omega$ we may approximate $\omega (\nn) \simeq T$, in which case we  recover the traditional formula 
$\Phi \simeq \frac{\Phi_E}{T}$ [Eq.~(\ref{Phi_classical})].
Thus, Eq.~(\ref{single:Phi_3}) recovers the expected result at high temperatures. In addition, it tends to a finite value as $T \to 0$.
Hence, as mentioned above, within the Wigner entropy formulation, both $\Pi$ and $\Phi$ remain well behaved in the limit $T\to 0$.

We have opted to derive Eqs.~(\ref{single:Pi}) and (\ref{single:Phi_3}) starting from the Wigner relative entropy, since this gives the most natural physical interpretation. 
In the supplemental material we provide two alternative derivations of these formulas. 
The first is through a simple algebraic manipulation, which makes no mention at all to the relative entropy or to the target state $W_{\text{eq}}$. 
It may therefore be useful in situations where one does not know the target state of the dissipator \emph{a priori}. 

The second method is  to  map the Fokker-Planck equation~(\ref{single:FP}) into a stochastic process in the complex plane. 
In this way, the total entropy production $\Sigma$ of a process may be defined as a functional of the stochastic forward and backward trajectories. 
The entropy production rate is then obtained by averaging the stochastic entropy over an infinitesimal time interval, $\langle \Sigma \rangle = \Pi \ud t$, where $\langle\cdot \rangle$ stands for the average over all stochastic paths. 
The interesting aspect of this approach is that, as we show,  $\Sigma$ satisfies an integral fluctuation theorem, which is the fundamental property expected of the entropy production.
This supports the interpretation of Eq.~(\ref{single:Pi}) as a valid entropy production rate.

%%%%%%%%%%%%%%%%%%%%%%%%%%%%%%%%%%%%%%
%
%
%						SQUEEZED BATH
%
%
%%%%%%%%%%%%%%%%%%%%%%%%%%%%%%%%%%%%%%

\emph{Squeezed bath - }We now generalize the above results to the case of a bosonic mode subject to a non-equilibrium broad-band squeezed bath.
This type of reservoir appears frequently  in quantum optics \cite{Pontin2016,Lotfipour2016,Clark2016,Castellanos-Beltran2008,Wu1986,Manzano2016}, whenever the squeeze radiation field is treated as an external bath. 
The bath is characterized  by a temperature $T$ (usually zero), a squeeze parameters $r e^{i \theta}$ and a central frequency $\omega_s$, related to the production of the squeezed field (usually by parametric down conversion). 

The dissipator of the squeezed bath may be written in terms of the squeezed operators $b_z = S(z) a S^\dagger(z)$, where $S(z) = e^{(z^* a^2 - z a^{\dagger 2})/2}$ and $z = r e^{i (\theta - 2 \omega_s t)}$.
With these definitions, the formula for the squeezed Lindblad super-operator becomes identical to Eq.~(\ref{single:D}), with $a$ replaced by $b_z$.
Due to this correspondence, all results obtained above for the thermal bath remain valid for the squeezed bath, provided the calculations are all carried out in terms of the operators $b_z$ instead of $a$. 
This allows us to readily write down the analogues of Eqs.~(\ref{single:Pi}) and (\ref{single:Phi_3}) as
\begin{equation}
\Pi=\frac{4}{\gamma(\nn)}  \int\ud^2 \beta \frac{|J_b(W)|^2}{W}, \Phi=\frac{\gamma(\langle b_z^\dagger b_z \rangle - \bar{n})}{\nn} ,
\label{squeeze:phi}
\end{equation}
where $J_b(W)$ is defined exactly as in Eq.~(\ref{single:J}), but with $\beta$ instead of $\alpha$. 
As $b_z$ and $a$ are related by a unitary transformation, the Jacobian of the transformation from $\beta$ to $\alpha$ is unity. 
Moreover, a straightforward calculation shows that 
\begin{equation}
J_b(W) {=} J(W)\cosh r  {+} [\gamma \alpha^*W {-} J^*(W)]e^{i (\theta- 2\omega_s t)} \sinh r  .
\end{equation}
With these transformations and $b_z = S(z) a S^\dagger(z)$, it is possible to express both $\Pi$ and $\Phi$ solely in terms of quantities linked to $a$ and $a^\dag$. 
%This allows us to write $\Pi$ and $\Phi$ in terms of $a$-quantities as 
%\begin{IEEEeqnarray}{rCl}
%\Pi&=& \frac{4/\gamma}{\nn}  \int\ud^2 \alpha \frac{|\cosh r J(W) + e^{i (\theta- 2\omega_s t)} \sinh r J^*(W)|^2}{W},
%\IEEEeqnarraynumspace		\\[0.2cm]
%\Phi&=& \frac{\gamma}{(\nn)^2} \bigg\{ (N+\frac{1}{2}) \langle a^\dagger a \rangle - \text{Re}[M_t^* \langle a a \rangle]+ \sinh^2 r - \bar{n}
%\bigg\}.\IEEEeqnarraynumspace
%\end{IEEEeqnarray}

To illustrate the new effects brought about by the squeezing of the bath, consider a cavity with frequency $\omega_c$ pumped by a radiation field with frequency $\omega_p$ and squeezed central frequency $\omega_s$. 
The Hamiltonian describing the cavity mode is 
\begin{equation}\label{spump:H}
H = \omega_c a^\dagger a + i (\mathcal{E} e^{-i \omega_p t} a^\dagger - \mathcal{E}^* e^{i \omega_p t} a),
\end{equation}
where ${|\cal{E}|} = \sqrt{2P\kappa/\hbar \omega_p}$, with $P$ being the pump laser power and $\kappa=\gamma/2$  the cavity amplitude decay rate. 
The contact with the squeezed reservoir is modeled exactly by the Lindblad super-operator Eq.~(\ref{single:D}), with $a\to b_z$, $\gamma = 2\kappa$ and $\bar{n} = 0$.
Due to the Gaussian nature of the problem, all calculations  are straightforward [cf. Supplemental Material]. 
Here we only emphasize the final result. 
First, the steady-state energy flux is given by 
\begin{equation}\label{squeeze_energy_flux}
\Phi_E = \left\langle \frac{\partial H}{\partial t} \right\rangle =  \frac{2\kappa \omega_p |\mathcal{E}|^2}{\kappa^2 + \Delta_{cp}^2},
\end{equation}
where $\Delta_{ij} = \omega_i - \omega_j$.
The heat current will thus be non-zero only in the presence of the pump.
Second, at the steady-state we have $\ud S/\ud t = 0$, so that $\Pi = \Phi$ and
\begin{equation}\label{spump:Phi}
\begin{aligned}
\Pi  &= \frac{2\kappa\Delta_{sc}^2}{\kappa^2 + \Delta_{sc}^2} {\sinh^2(2r)} + \frac{4\kappa|\mathcal{E}|^2}{\kappa^2 + \Delta_{cp}^2} \cosh(2r)\\
&+4\kappa \text{Re}\bigg[ \frac{\mathcal{E}^2e^{-i(2 \Delta_{ps} t + \theta)}}{(\kappa+ i \Delta_{cp})^2} \bigg] \sinh(2r). %\nonumber
\end{aligned}
\end{equation}
If $\omega_p \neq \omega_s$ and in a time-averaged picture, the last term will oscillate in time with zero average and may thus be neglected.
In the limit of zero squeezing ($r\to 0$) only the second term survives and we find that $\Phi = \Phi_E/2$. 
\begin{figure}
\centering
\includegraphics[width=0.6\columnwidth]{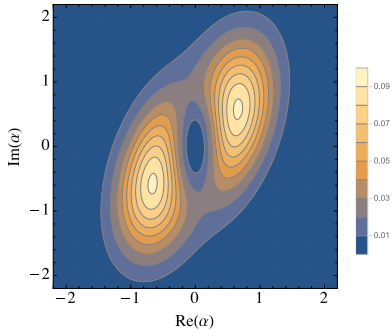}
\caption{\label{fig:jb}  $|J_b|^2/W$ as a function of $\alpha$, computed using Eq.~(\ref{Jb}) with $\Delta_{cs}/\kappa = 0.9$, $r = 0.5$ and $\theta- 2 \omega_s t = 0.0$. }
\end{figure}
%The  terms which depend on  $\mathcal{E}$ represent the entropy production/flux due to the external pump. 
The important part of Eq.~(\ref{spump:Phi}), however, is  the first term, which would still be present even for no pumping ($\mathcal{E} = 0$). 
This term reflects the contribution coming from the non-equilibrium nature of the bath (encompassed by the degree of squeezing), and the one resulting from the  mismatch between the central frequency $\omega_s$ of the broad-band squeezed bath and the natural frequency $\omega_c$ of the cavity (which induces off-resonant exchanges of excitations between the cavity and the bath that are not present in the resonant case).
We interpret this term as a signature of a (irreversible) non-equilibrium steady-state that will occur even in the absence of a pump. 

It is remarkable that this information is not present in the energy flux rate Eq.~(\ref{squeeze_energy_flux}), but only in the entropy production.
This thus provides a clear exception to the usual assumption that non-equilibrium steady-states always have an associated energy current. 
In addition, our formulation reveals a genuinely quantum effect, and one that in principle could be measured experimentally. Similar counter-intuitive results have been reported for the efficiency of quantum Carnot cycles under squeezed reservoirs \cite{Ronagel2014}. 
We can also analyze this effect from the view-point of the irreversible current $J_b(W)$ appearing in Eq.~(\ref{squeeze:phi}). 
Using the results detailed in the Supplemental Material, one may readily show that for $\mathcal{E} =0$ 
\begin{equation}\label{Jb}
\frac{|J_b(W)|^2}{W^2} = \frac{\kappa^2 \Delta_{sc}^2 \sinh^2(2r)}{\kappa^2 + \Delta_{cs}^2 \cosh^2(2r)} |\beta|^2
\end{equation}
where $\beta = \alpha \cosh r + \alpha^* e^{i (\theta- 2 \omega_s t)} \sinh r$.
Thus, the magnitude of the current will be zero when either $\Delta_{sc}=0$ or $r =0$.
Fig.~\ref{fig:jb} shows $|J_b|^2/W$ against  $\alpha$.

%%%%%%%%%%%%%%%%%%%%%%%%%%%%%%%%%%%%%%
%
%
%			DEPHASING BATH
%
%
%%%%%%%%%%%%%%%%%%%%%%%%%%%%%%%%%%%%%%

\emph{Dephasing bath - }Finally, we turn to the problem of a dephasing bath, characterized by the Lindblad super-operator 
\begin{equation}\label{gen:dephasing:Drho}
\mathcal{D}_\text{deph}(\rho) = \lambda \bigg[ a^\dagger a \rho a^\dagger a - \frac{1}{2} \{ (a^\dagger a)^2,\rho\}\bigg].
\end{equation}
The action of the environment is to suppress quantum coherences without the exchange of energy with the system, so that $\Phi_E = 0$.
The corresponding operator in Wigner space reads 
$\mathcal{D}_\text{deph}(W) = \partial_\alpha I(W) + \partial_{\alpha^*} I^*(W)$
where
%\begin{equation}\label{gen:deph:I}
$I(W) = {\lambda}\alpha \left[ \alpha^* \partial_{\alpha^*} W - \alpha\partial_\alpha W\right]/2$. 
%\end{equation}
The target state of this dynamics is not unique, as any Gibbs thermal state will be a target state.
Using Eq.~(\ref{single:Pi:relative}), we find 
\begin{equation}\label{gen:deph:Pi}
\frac{\ud S}{\ud t} \bigg|_\text{deph} = \Pi_\text{deph} = \frac{2}{\lambda} \int\frac{\ud^2\alpha}{W} \frac{|I(W)|^2}{|\alpha|^2}.
\end{equation}
Clearly, for such a dephasing bath the entropy flux $\Phi$ is null, which agrees intuitively with the idea that the energy flux will also be zero, and demonstrate the suitable nature of the approach that we have proposed.

\emph{Conclusions - }We have addressed the difficulty of studying irreversibility in general quantum process incorporating an out-of-equilibrium environment. While relevant, experimentally and technologically, in a number of physical situations (especially in solid-state quantum information processing), the successful addressing of this problem has so far been hindered by the lack of formal tools suited to encompass the complexity of the effects arising from the environment. We have put forward a new, alternative picture for irreversible entropy production based on the use of R\'enyi-2 entropy, which is able to address the open-system dynamics of a quantum system in contact with non-equilibrium reservoirs in a successful way. 
Three independent methods of obtaining the entropy production rate were provided, which serves to corroborate the generality of our approach.
Whether it is possible to generalize this theory to  arbitrary R\'enyi-$\alpha$ entropies remains an open question. 
The illustrations that we have discussed, including squeezed and dephasing baths, show both the potential of the proposed approach and the breath of physically relevant situation that it is able to address. 
We have opted to focus our approach on a single bosonic mode. 
The generalization to a multi-mode process is straightforward and will be the subject of a future publication. 

{\it Acknowledgements.--}G. T. L. would like to acknowledge the S\~ao Paulo Research Foundation, under grant number
2016/08721-7. 
J. P. Santos would like to acknowledge the financial support from the CAPES (PNPD program) for the postdoctoral grant. MP acknowledges support from the EU Collaborative project TherMiQ (grant agreement 618074), the Julian Schwinger Foundation (grant number JSF-14-7-0000), the DfE-SFI Investigator Programme (grant 15/IA/2864) and the Royal Society Newton Mobility Grant NI160057. This work was partially supported by the COST Action MP1209.

\bibliography{/Users/gtlandi/Documents/library}

%%%%%%%%%% Merge with supplemental materials %%%%%%%%%%
\pagebreak
\widetext
\begin{center}
\textbf{\large Supplemental Materials: The Wigner entropy production rate}
\end{center}
%%%%%%%%%% Merge with supplemental materials %%%%%%%%%%
%%%%%%%%%% Prefix a "S" to all equations, figures, tables and reset the counter %%%%%%%%%%
\setcounter{equation}{0}
\setcounter{figure}{0}
\setcounter{table}{0}
\setcounter{page}{1}
\makeatletter
\renewcommand{\theequation}{S\arabic{equation}}
\renewcommand{\thefigure}{S\arabic{figure}}
\renewcommand{\bibnumfmt}[1]{[S#1]}
\renewcommand{\citenumfont}[1]{S#1}
%%%%%%%%%% Prefix a "S" to all equations, figures, tables and reset the counter %%%%%%%%%%

In this Supplemental Material we show two alternative routes for deriving the formulas for the Wigner entropy production rate and Wigner entropy flux rate, that form the main results of this paper. 
The first (Sec. 1) is by means of a purely algebraic method and the second (Secs. 2 and 3) is based on averaging stochastic trajectories in the complex plane. 
In Sec. 4 we also give further details on the calculation of the steady-state of a pump cavity in a squeezed reservoir,  in particular Eq.~(21) of the main text.

%%%%%%%%%%%%%%%%%%%%%%%%%%%%%%%%%%%%%%%%%%%%%%%%%%%%%%
\section{Algebraic procedure}

In order to be more general, we will consider here the case of a squeezed bath, which already encompass a normal heat bath as a limiting case. 
Moreover, in order to provide an alternative view of the calculations done in the main text, we will work here in the $a$ representation. 
The dissipator then reads
\begin{IEEEeqnarray}{rCl}\label{gen:squeeze:D2}
\mathcal{D}_z(\rho) &=& \gamma(N+1)\bigg[ a \rho a^\dagger - \frac{1}{2} \{a^\dagger a, \rho\}\bigg] \\[0.2cm]
&&+ \gamma  N \bigg[ a^\dagger \rho a - \frac{1}{2} \{ a a^\dagger, \rho\}\bigg]\nonumber \\[0.2cm]
&& - \gamma M_t  \bigg[ a^\dagger \rho a^\dagger - \frac{1}{2} \{ a^\dagger a^\dagger, \rho\}\bigg]\nonumber \\[0.2cm]
&& - \gamma M_t^*   \bigg[ a \rho a - \frac{1}{2} \{ a a, \rho\}\bigg]\nonumber 
\end{IEEEeqnarray}
where
\begin{IEEEeqnarray}{rCl}\label{gen:squeeze:N}
N+\nicefrac{1}{2} &=& (\nn) \cosh 2r 	\nonumber \\[-0.2cm]
\label{gen:squeeze:NM}\\
M_t &=& -(\nn) e^{i (\theta -2\omega_s t)} \sinh 2r\nonumber
\end{IEEEeqnarray}
The  dissipator~(\ref{gen:squeeze:D2}) in Wigner space becomes
\begin{equation}\label{dissipador_Jz}
\mathcal{D}_z(W) = \partial_\alpha J_z(W) + \partial_{\alpha^*} J_z^*(W)
\end{equation}
where
\begin{equation}\label{gen:squeeze:Jz}
J_z(W) = \frac{\gamma}{2} \bigg[ \alpha W + (N+\nicefrac{1}{2}) \partial_{\alpha^*} W + M_t  \partial_\alpha W\bigg]
\end{equation}
In the limit $r\to0$ we recover the usual thermal bath dissipator.

We now study the rate of change of the entropy, 
\begin{equation}
\frac{\ud S}{\ud t} = - \int \ud^2\alpha (\partial_t W) \ln W
\end{equation}
Inserting the full Fokker-Planck equation for $\partial_t W$ one finds that the unitary part does not contribute to $\ud S/\ud t$. 
We are then left with
\begin{IEEEeqnarray*}{rCl}
\frac{\ud S}{\ud t} &=& - \int \ud^2\alpha \mathcal{D}_z(W) \ln W		\\[0.2cm]
&=& \int \frac{\ud^2\alpha}{W} \bigg[ J_z(W) \partial_\alpha W + J_z^*(W) \partial_{\alpha^*} W\bigg]
\end{IEEEeqnarray*}
where in the last line we integrated by parts. 
It is more convenient to write this as a dot product
\begin{equation}\label{gen:squeeze:Stmp1}
\frac{\ud S}{\ud t} = \int\frac{\ud^2 \alpha}{W} \begin{pmatrix} J_z^* & J_z \end{pmatrix}
\begin{pmatrix} \partial_{\alpha^*} W \\[0.2cm] \partial_{\alpha} W \end{pmatrix}
\end{equation}

From Eq.~(\ref{gen:squeeze:Jz}) we have 
\begin{equation}
A \begin{pmatrix} \partial_{\alpha^*} W \\[0.2cm] \partial_{\alpha} W \end{pmatrix} = \frac{2}{\gamma} \begin{pmatrix} J_z \\[0.2cm] J_z^* \end{pmatrix} -
\begin{pmatrix} \alpha \\[0.2cm] \alpha^* \end{pmatrix} W
\end{equation}
where
\begin{equation}
A = \begin{pmatrix}
N+\nicefrac{1}{2}	& M_t \\[0.2cm] M_t^* & N + \nicefrac{1}{2} \end{pmatrix}
\end{equation}
%is the {\color{cyan} Bla-bla-bla covariance matrix of squeezed states [ref]}.
The matrix $A$ turns out to be the covariance matrix for a system in a squeezed Gibbs state.

Solving for $\partial_{\alpha} W$ and $\partial_{\alpha^*} W$ and substituting the result in Eq.~(\ref{gen:squeeze:Stmp1}) allows us to separate $\ud S/\ud t$ as
\begin{equation}
\frac{\ud S}{\ud t} = \Pi - \Phi
\end{equation}
where
\begin{IEEEeqnarray}{rCl}
\label{gen:squeeze:Pi1}
\Pi &=& \frac{2}{\gamma} \int\frac{\ud^2\alpha}{W} \begin{pmatrix} J_z^* & J_z \end{pmatrix} A^{-1} \begin{pmatrix} J_z \\[0.2cm] J_z^* \end{pmatrix}	\\[0.2cm]
\label{gen:squeeze:Phi1}
\Phi &=&  \int\ud^2\alpha  \begin{pmatrix} J_z^* & J_z \end{pmatrix} A^{-1} \begin{pmatrix} \alpha \\[0.2cm] \alpha^* \end{pmatrix}
\end{IEEEeqnarray}
The formula for $\Pi$, in particular, is now written as a quadratic form. 
Its positivity is hence ensured by the fact that the matrix $A$ is positive definite [its eigenvalues are the variances of the squeezed quadratures $(\nn) e^{\pm 2r}$].

It is also possible to simplify these results further by substituting for $A^{-1}$. 
We then get, after some manipulations,
\begin{IEEEeqnarray}{rCl}
\label{gen:squeeze:Pi1}
\Pi &=& \frac{4/\gamma}{\nn} \int\frac{\ud^2\alpha}{W} |J_z\cosh r + J_z^* e^{i ( \theta-2\omega_s t)} \sinh r|^2	
\IEEEeqnarraynumspace
\\[0.2cm]
\label{gen:squeeze:Phi1}
%\Phi &=& \frac{\gamma}{\nn} \bigg[  \langle a^\dagger a \rangle \cosh2r + \sinh^2 r - \bar{n} +\\[0.2cm]
%&&- \bigg(\frac{\langle a a \rangle}{2} e^{-i \theta} e^{-i 2(\omega-\omega_s)t}+ \frac{\langle a^\dagger a^\dagger \rangle}{2} e^{i \theta} e^{i 2(\omega-\omega_s)t}\bigg) \sinh2r \bigg]\nonumber
%\\[0.2cm]
\Phi 
%\frac{\gamma}{\nn} \bigg[ \langle b_z^\dagger b_z \rangle - \bar{n}\bigg]	\\[0.2cm]
% &=& \frac{\gamma}{(\nn)^2} \bigg\{ (N+\nicefrac{1}{2}) \langle a^\dagger a \rangle - \text{Re}[M_t^* \langle a a \rangle]+ \sinh^2 r - \bar{n} Eu acho que essa relação está errada
&=&\frac{\gamma}{\nn} \bigg\{ \cosh(2r)\langle a^\dagger a \rangle  -\bar{n} +\sinh^2(r)- 
\frac{\text{Re}[M_t^* \langle a a \rangle]}{\nn}
\bigg\}
\end{IEEEeqnarray}
We will show next that these results are actually identical  to the formulas in Eq.~(17) of the main manuscript.

We can relate the current $J_z$ to the current $J_b$ used in the main text:
\begin{equation}
J_b(W) = \frac{\gamma}{2} \bigg[ \beta W + (\bar{n}+\nicefrac{1}{2}) \partial_{\beta^*} W\bigg]
\end{equation}
To do so we use the change of variables
\begin{IEEEeqnarray*}{rCl}
\beta&=&\alpha\cosh r + e^{+i(\theta-2\omega_s t)}\alpha^*\sinh r.
\end{IEEEeqnarray*}
In this case, we can rewrite the Eq.~(\ref{dissipador_Jz}) as 
\begin{IEEEeqnarray*}{rCl}
\mathcal{D}_z(W) &=& \partial_\beta [ J_z \cosh r  + J_z^*\sinh r e^{+i(\theta-2\omega_st) }]  \\[0.2cm]
                 &+& \partial_{\beta^*}[J_z^*\cosh r  + J_z\sinh r e^{-i(\theta-2\omega_st) }] \\[0.2cm]
                 &=& \partial_\beta J_b  + \partial_{\beta^*} J_b^*
\end{IEEEeqnarray*}
where
\begin{equation} \label{eq:Jb}
J_b = J_z \cosh r  + J_z^*\sinh r e^{+i(\theta-2\omega_st)}
\end{equation}
Now, one can readily rewrite the entropy production (\ref{gen:squeeze:Pi1}) as
\begin{equation}
\Pi = \frac{4/\gamma}{\nn} \int\frac{\ud^2\beta}{W} |J_b|^2
\end{equation}
This is exactly Eq.~(17) of the main text.
Thus, we have just derived the formula for the entropy production without any mention of the Wigner relative entropy, or the target state of the squeezed dissipator. 
Instead, the entire demonstration is based on an algebraic separation of $\ud S/\ud t$ into two convenient terms, one of which is always non-negative. 
While this derivation may lack a more  physical justification, it may be useful in situations where the target state is not easily known.

Finally, we can also relate the current $J_b$ to the current $J$, which is its $\alpha$-representation, defined in Eq.~(10) of the main text:
\begin{equation}
J(W) = \frac{\gamma}{2} \bigg[ \alpha W + (\bar{n}+\nicefrac{1}{2}) \partial_{\alpha^*} W\bigg]
\end{equation}
In order to do that we substitute the expression for $J_z$ in (\ref{eq:Jb}).
After some manipulations we obtain
\begin{IEEEeqnarray*}{rCl}
J_b &=& 
\cosh r \bigg[ \frac{\gamma}{2} \bigg( 
\alpha W + (\bar{n}+\nicefrac{1}{2}) \partial_{\alpha^*} W \bigg)\bigg]
\\[0.2cm]
&-& 
\sinh r e^{+i(\theta-2\omega_st)} \bigg[ \frac{\gamma}{2} \bigg( 
\alpha^* W + (\bar{n}+\nicefrac{1}{2}) \partial_{\alpha} W \bigg)\bigg]
\\[0.2cm]
&+&
\cosh r \bigg[ \frac{\gamma}{2} \bigg(
[(N+\nicefrac{1}{2})-(\bar{n}+\nicefrac{1}{2})]\partial_{\alpha^*} W+ M_t\partial_{\alpha} W
\bigg)\bigg]
\\[0.2cm]
&+&
\sinh r e^{+i(\theta-2\omega_st)} \bigg[ 
\frac{\gamma}{2} \bigg(
2\alpha^*W+[(N+\nicefrac{1}{2})+(\bar{n}+\nicefrac{1}{2})]\partial_{\alpha} W+ M_t^*\partial_{\alpha^*} W
\bigg)\bigg]
\end{IEEEeqnarray*}
Using the results 
\begin{IEEEeqnarray*}{rCl}
(N+\nicefrac{1}{2})-(\bar{n}+\nicefrac{1}{2}) &=& 
-\frac{\sinh r}{\cosh r} M_t^* e^{i(\omega-2\omega_st)}
\\[0.2cm]
(N+\nicefrac{1}{2})+(\bar{n}+\nicefrac{1}{2}) &=& 
-\frac{\cosh r}{\sinh r} M_t e^{-i(\omega-2\omega_st)}
\end{IEEEeqnarray*}
We may then write $J_b$ as 
\begin{equation}
J_b = \cosh r J + e^{i(\theta-2\omega_st)} \sinh r [\gamma\alpha^*W-J^*]
\end{equation}
which is Eq.~(18) of the main text.

%%%%%%%%%%%%%%%%%%%%%%%%%%%%%%%%%%%%%%%%%%%%%%%%%%%%%%
\section{Stochastic trajectories on the complex plane}

Another way of deriving the formulas for $\Pi$ and $\Phi$ is by analyzing quantum trajectories in the complex plane. 
For simplicity, we work only with the thermal heat bath, for which the Fokker-Planck equation reads 
 \begin{equation}\label{single:FP}
{\partial_t W} = - i \omega\bigg[ \partial_{\alpha^*} (\alpha^* W) - \partial_\alpha (\alpha W)\bigg] + \mathcal{D}(W),
\end{equation}
where
\begin{equation}\label{single:DW_J}
\mathcal{D}(W) = \partial_\alpha J(W) + \partial_{\alpha^*} J^*(W),
\end{equation}
and
\begin{equation}\label{single:J}
J(W) = \frac{\gamma}{2} \bigg[ \alpha W + (\overline{n}+1/2) \partial_{\alpha^*} W\bigg].
\end{equation}
This Fokker-Planck equation can be modeled by a complex stochastic  variable $A(t)$ satisfying the Langevin equation:
\begin{equation}\label{single:langevin}
\frac{\ud A}{\ud t} = -i \omega A - \frac{\gamma}{2}A + \sqrt{\gamma (\nn)} \xi(t)
\end{equation}
where $\xi(t)$ is a complex Gaussian white noise:
\begin{equation}
\langle \xi(t) \xi(t') \rangle = 0, \qquad \langle \xi(t) \xi^*(t') \rangle = \delta(t-t')
\end{equation}

Now suppose that the process takes place between a time $t = 0$ and $t = \tau$ and let $\alpha(t)$ denote a possible stochastic trajectory of $A(t)$. 
We denote by $\mathcal{P}[\alpha(t)]$ the  probability of observing the path $\alpha(t)$. 
In addition we define the time-reversed trajectory $\alpha^*(\tau-t)$ and we let $\mathcal{P}_R[\alpha^*(\tau-t)]$ denote the corresponding probability of observing the time-reversed trajectory. 
We then define the entropy produced in a given stochastic trajectory $\alpha(t)$ as 
\begin{equation}\label{single:Sigma}
\Sigma[\alpha(t)] = \ln \frac{\mathcal{P}[\alpha(t)]}{\mathcal{P}_R[\alpha^*(\tau-t)]}
\end{equation}
The entropy produced is the ratio of the forward and backward probabilities. 
When a given path is reversible, the two probabilities coincide and  no entropy is produced. 

The strongest argument corroborating the correctness of interpreting Eq.~(\ref{single:Sigma}) as an entropy production is that it satisfies a fluctuation theorem: 
\begin{IEEEeqnarray*}{rCl}
\langle e^{- \Sigma[A(t)]}\rangle &=& \int D\alpha(t) \mathcal{P}[\alpha(t)] e^{-\ln \mathcal{P}[\alpha(t)]/\mathcal{P}_R[\alpha^*(\tau-t)]}	\\[0.2cm]
&=& \int D\alpha^*(\tau-t) \mathcal{P}_R[\alpha^*(\tau-t)]\\[0.2cm]
&=& 1
\end{IEEEeqnarray*}
Here $D\alpha(t)$ is the integration measure of the path integral and in the second line, use was made of the fact that the Jacobian determinant of the transformation from $\alpha(t)$ to $\alpha^*(\tau-t)$ is unity.

To find the entropy production rate we consider an infinitesimal stochastic path taking place between times $t$ and $t + \ud t$, and let $\ud \Sigma[A(t)]$ denote the entropy produced in it. 
We will now show that it is possible to obtain the formula for the entropy production rate [Eq.~(14) of the main text] from
\begin{equation}\label{single:stochastic_Pi1}
\Pi = \frac{\langle \ud \Sigma[A(t)] \rangle}{\ud t}
\end{equation}
To carry out this demonstration we first note that the path $\alpha(t)$ in an infinitesimal  process is reduced to two complex numbers, $\alpha = \alpha(t)$ and $\alpha' = \alpha(t+\ud t)$. 
The probability $\mathcal{P}[\alpha(t)]$ for the path may then be written as 
\begin{equation}\label{single:stochastic_P}
\mathcal{P}[\alpha', \alpha] = \mathcal{K}_{\ud t} (\alpha' | \alpha) W(\alpha,t)
\end{equation}
where $W(\alpha,t)$ is a shorthand notation for $W(\alpha, \alpha^*,t)$,  which is the Wigner function at time $t$.
Moreover, $\mathcal{K}_{t}$ is the  propagator of the system, defined from 
\begin{equation}\label{single:stochastic_Kdef}
W(\alpha',t) = \int\ud^2 \lambda \mathcal{K}_{t-t_0}(\alpha' | \alpha) W(\alpha,t_0)
\end{equation}
As shown in the next section, the infinitesimal version $\mathcal{K}_{\ud t}$ of the propagator, for the  Fokker-Planck equation~(\ref{single:FP}), is 
\begin{equation}\label{single:stochastic_prop}
\mathcal{K}_{\ud t}(\alpha'|\alpha) = \frac{e^{\gamma \ud t}}{ \pi \gamma (\nn)} 
\exp\bigg\{ - \frac{| \alpha'[ 1 +\ud t(i \omega + \gamma/2)]  - \alpha |^2}{\gamma \ud t (\nn)}\bigg\}
\end{equation}

Eq.~(\ref{single:stochastic_P}) gives the probability for the forward path. 
The corresponding probability for the reversed path is 
\begin{equation}\label{single:stochastic_PR}
\mathcal{P}_R[\alpha^*, \alpha'^*] = \mathcal{K}_{\ud t} (\alpha^*| \alpha'^*) W(\alpha'^*, t+ \ud t)
\end{equation}
From Eq.~(\ref{single:Sigma}) we then find that the entropy produced in the infinitesimal trajectory will be 
\begin{IEEEeqnarray}{rCl}\label{single:stochastic_dSigma_def}
\ud \Sigma[A(t)] &=& \ln \frac{W(A(t),t)}{W(A^*(t+\ud t), t+ \ud t)}  \\[0.2cm]
&&+ \ln \frac{ \mathcal{K}_{\ud t}(A(t+\ud t)| A(t))}{ \mathcal{K}_{\ud t}(A^*(t) | A^*(t + \ud t))}\nonumber
\end{IEEEeqnarray}
We must now average this stochastic number over the distribution $\mathcal{P}[\alpha',\alpha]$ of Eq.~(\ref{single:stochastic_P}). 
Since we do not know what is the Wigner function $W(\alpha,t)$, we may only carry out the integral over $\alpha'$ and leave the results as an average over $W(\alpha,t)$. 
Moreover, since  the propagator is Gaussian, all integrals may be computed without difficulty. 

We start with the second term in Eq.~(\ref{single:stochastic_dSigma_def}). 
From Eqs.~(\ref{single:stochastic_P}) and (\ref{single:stochastic_PR}) we find that 
\begin{equation}\label{single:stochastic_2nd_term}
\ln \frac{ \mathcal{K}_{\ud t}(\alpha'|\alpha)}{ \mathcal{K}_{\ud t}(\alpha^*|\alpha'^*)} = \frac{|\alpha|^2 - |\alpha'|^2}{\nn} + \mathcal{O}(\ud t)^2
\end{equation}
However, from Eq.~(\ref{single:stochastic_prop}) it follows that 
\begin{IEEEeqnarray*}{rCl}
\int\ud^2 \alpha' \; (|\alpha|^2 - |\alpha'|^2 ) \mathcal{K}_{\ud t} (\alpha'|\alpha)
= \\[0.2cm]
\gamma \ud t  \bigg[ |\alpha|^2 - (\nn)\bigg] 
\end{IEEEeqnarray*}
Hence,  upon averaging  Eq.~(\ref{single:stochastic_2nd_term}) we get, to first order in $\ud t$,
\begin{IEEEeqnarray}{rCl}
\label{single:stoch:ave1}
\bigg\langle \ln \frac{ \mathcal{K}_{\ud t}(A(t+\ud t)| A(t))}{ \mathcal{K}_{\ud t}(A^*(t) | A^*(t + \ud t))} \bigg\rangle = \\[0.2cm] \nonumber
\frac{\gamma\ud t}{\nn} \int\ud^2 \alpha \bigg[ |\alpha|^2 - (\nn)\bigg] W(\alpha,t)
\end{IEEEeqnarray}

Next we turn to the first term in Eq.~(\ref{single:stochastic_dSigma_def}). 
To compute it we must use It\^o's lemma. First we write 
\begin{equation}\label{single:stochastic_lnW}
\ln \frac{W(A(t),t)}{W(A^*(t+\ud t), t+ \ud t)} = - \ud \bigg[ \ln  W(A(t),t) \bigg]
\end{equation}
According to It\^o's lemma, if $f(A)$ is an arbitrary function of $A$, then 
\begin{equation}
\ud f = \frac{\partial f}{\partial A} \ud A + \frac{\partial f}{\partial A^*} \ud A^* + \bigg[  \frac{\partial f}{\partial t}  +\gamma(\nn) \frac{\partial^2 f}{\partial A\partial A^*} \bigg] \ud t
\end{equation}
Applying this to Eq.~(\ref{single:stochastic_lnW}) leads to 
\begin{IEEEeqnarray}{rCl}\label{single:stochastic_Ito}
&& \ud \bigg[ \ln  W(A(t),t) \bigg] =   \frac{1}{W} \frac{\partial W}{\partial A} \ud A +  \frac{1}{W} \frac{\partial W}{\partial A^*} \ud A^*
 \\[0.2cm]
&& +
 \bigg[ \frac{1}{W}\frac{\partial W}{\partial t} 
- \frac{\gamma (\nn)}{W} \bigg( \frac{\partial^2 W}{\partial A \partial A^*} - \frac{1}{W} \frac{\partial W}{\partial A} \frac{\partial W}{\partial A^*}\bigg) \bigg] \ud t \nonumber 
\end{IEEEeqnarray}

When averaging this result we note $A(t+\ud t)$ appears only in $\ud A$ and $\ud A^*$, with all other terms being functions only of $A(t)$ and $A^*(t)$. 
Consequently, when averaging, we may carry out the integration over $\alpha'$ and eliminate the contribution of the propagator:
\begin{IEEEeqnarray*}{rCl}
\langle \ud A \rangle &=& \langle A(t+ \ud t) - A(t) \rangle \\[0.2cm]
&=& \int \ud^2 \alpha'\ud^2 \alpha\;  (\alpha'-\alpha) K_{\ud t}(\alpha'|\alpha)W(\alpha,t) \\[0.2cm]
&=& - \left(\frac{\gamma}{2} + i \omega\right) \ud t \int \ud^2 \alpha \; \alpha\;  W(\alpha,t)
\end{IEEEeqnarray*}
Most of the remaining integrals over $\alpha$ in Eq.~(\ref{single:stochastic_Ito}) will turn out to be zero. 
All that remains is
%\begin{IEEEeqnarray*}{rCl}
%\int \ud^2 \alpha\; \alpha \;  \partial W  &=& - 1	\\[0.2cm]
%\int \ud^2 \alpha \frac{(\partial W) (\partial^* W)}{W} &=& \int\ud^2 \alpha W |\partial \ln W|^2 
%\end{IEEEeqnarray*}
%Thus, we conclude that 
\begin{IEEEeqnarray}{rCl}
\nonumber
- \bigg\langle\ud \bigg[ \ln  W(A(t),t) \bigg] \bigg \rangle
&=&
\ud t \int\ud^2 \alpha \frac{\gamma}{\nn} \bigg[ -  (\nn)
\\[0.2cm]
&& + (\nn)^2 |\partial \ln W|^2 \bigg] W
\label{single:stoch:ave2}
\end{IEEEeqnarray}

Combining Eqs.~(\ref{single:stoch:ave1}) and (\ref{single:stoch:ave2}) then finally yields 
\begin{IEEEeqnarray}{rCl}\nonumber
\langle \ud \Sigma[A(t)] \rangle &=& \frac{\gamma \ud t}{\nn} \int\ud^2\alpha\; \bigg[ |\alpha|^2 - 2 (\nn) \\[0.2cm]
&&+  (\nn)^2 |\partial \ln W|^2 \bigg] W
\end{IEEEeqnarray}
Returning now to the definition~(\ref{single:J}), we find that the quantity inside this integral is proportional to $|J(W)|^2$ so that this result may be written as 
\begin{equation}
\langle \ud \Sigma[A(t)] \rangle = \frac{4\ud t/\gamma }{\nn} \int\frac{\ud^2\alpha}{W} |J(W)|^2
\end{equation}
Finally, dividing by $\ud t$ we obtain 
\begin{equation}
\Pi =  \frac{4/\gamma }{\nn} \int\frac{\ud^2\alpha}{W} |J(W)|^2
\end{equation}
which is Eq.~(14) of the main manuscript. 

As a side comment we note that the dephasing Fokker-Planck equation [Eq.~(24) of the main manuscript] may also be modeled as a stochastic process, but with a Langevin equation 
\begin{equation}\label{gen:deph:langevin}
\frac{\ud A}{\ud t} = -i \omega A  - \lambda A + i \sqrt{2 \lambda} A^* \xi(t)
\end{equation}
where $\xi(t)$ is a real Gaussian white noise satisfying $\langle \xi(t) \xi(t') \rangle = \delta(t-t')$. 
Discretizing time one may verify that 
\begin{equation}
\langle |A(t+\delta t)|^2 \rangle = \langle |A(t)|^2 \rangle + \mathcal{O}(\delta t)^2
\end{equation}
Thus, Eq.~(\ref{gen:deph:langevin}) describes a stochastic process which preserves the magnitude of the coherent state, changing only its phase. 
However, we see that in this case the noise is multiplicative and finding the corresponding short-time propagator is a much more difficult task.

\section{\label{app:prop}Infinitesimal propagator}

Lastly,  we derive a formula for the infinitesimal propagator $\mathcal{K}_{\ud t}$ defined in Eq.~(\ref{single:stochastic_Kdef}). 
The full propagator $\mathcal{K}_{t-t_0}(\alpha|\beta)$ is the Green function of the Fokker-Planck equation~(\ref{single:FP}):
\begin{equation}
\bigg[ \frac{\partial}{\partial t} - \mathcal{L}\bigg] \mathcal{K} = \delta(\alpha - \beta ) \delta(t - t_0)
\end{equation}
where $\mathcal{L}$ is the differential operator in the entire right-hand side of Eq.~(\ref{single:FP}), with all derivatives acting on $\alpha$. 
For short times $t - t_0 = \ud t$ we may write the propagator as 
\begin{equation}\label{single:stochastic:propagator_1}
\mathcal{K}_{\ud t} (\alpha| \beta) = e^{\ud t \mathcal{L}} \delta(\alpha- \beta) \simeq \bigg[ 1 + \ud t \mathcal{L} + \ldots \bigg] \delta(\alpha-\beta)
\end{equation}
Now we introduce the integral representation of the delta-function
\begin{equation}
\delta(\alpha) = \int \frac{\ud^2\lambda}{\pi^2} e^{-\lambda \alpha^* + \lambda^* \alpha}
\end{equation}
On the one hand, 
\begin{IEEEeqnarray*}{rCl}
\mathcal{L}(e^{-\lambda(\alpha^* - \beta^*) + \lambda^* (\alpha - \beta)}) &=& \bigg\{ \gamma - \gamma (\nn) |\lambda|^2 \\[0.2cm]
&&+ \left(\frac{\gamma}{2} + i \omega\right) \alpha \lambda^*- \left(\frac{\gamma}{2} - i \omega\right) \alpha^* \lambda\bigg\} e^{-\lambda(\alpha^* - \beta^*) + \lambda^* (\alpha - \beta)}
\end{IEEEeqnarray*}
But since $\ud t$ is infinitesimal, we may use this to approximate
\begin{IEEEeqnarray*}{rCl}
e^{\ud t \mathcal{L}}(e^{-\lambda(\alpha^* - \beta^*) + \lambda^* (\alpha - \beta)}) &=&\exp \bigg\{ \gamma \ud t - \gamma \ud t (\nn) |\lambda|^2 \\[0.2cm]
&&+ \lambda^* \bigg[ \alpha - \beta + \alpha \ud t (i\omega + \gamma/2)\bigg] 
- \lambda \bigg[ \alpha^* - \beta^* + \alpha^* \ud t (-i\omega + \gamma/2)\bigg]\bigg\}
\end{IEEEeqnarray*}
Inserting this result in Eq.~(\ref{single:stochastic:propagator_1}) will lead to a Gaussian integral, whose result is precisely Eq.~(\ref{single:stochastic_prop}).

%%%%%%%%%%%%%%%%%%%%%%%%%%%%%%%%%%%%%%%%%%%%%%%%%%%%%%
\section{Steady-state of a pumped cavity in a squeezed reservoir}

We consider here in more detail the problem of a pumped cavity in a squeezed reservoir.
The Hamiltonian of the system is given by 
\begin{equation}\label{spump:H}
H = \omega_c a^\dagger a + i (\mathcal{E} e^{-i \omega_p t} a^\dagger - \mathcal{E}^* e^{i \omega_p t} a)
\end{equation}
where ${|\cal{E}|} = \sqrt{2P\kappa/\hbar \omega_p}$, with $P$ being the pump laser power and $\kappa=\gamma/2$  the cavity amplitude decay rate. 
The squeezed reservoir can be modeled by Eq.~(\ref{gen:squeeze:D2}) with $\gamma=2\kappa$ and $\bar{n}=0$.

We begin by moving to a frame rotating at the pump frequency $\omega_p$. 
The effective Hamiltonian then changes to 
\begin{equation}\label{spump:H2}
H = \Delta_{cp} a^\dagger a + i (\mathcal{E} a^\dagger - \mathcal{E}^*  a)
\end{equation}
where $\Delta_{i,j} = \omega_i - \omega_j$ is the cavity detuning. 
The dissipator Eq.~(\ref{gen:squeeze:D2}) maintains the same form, except that the time-dependence of $M_t$ changes to 
$e^{-2i \Delta_{sp} t}$ instead of $e^{-2 i \omega_s t}$.

From the resulting master equation one may find dynamical equations for all observables of interest. 
The most important equations are:
\begin{IEEEeqnarray*}{rCl}
\frac{\ud \langle a \rangle}{\ud t} &=& \mathcal{E}  - (\kappa + i \Delta_{cp}) \langle a \rangle	\\[0.2cm]
\frac{\ud \langle a^\dagger a \rangle}{\ud t} &=& \mathcal{E} \langle a^\dagger \rangle + \mathcal{E}^* \langle a \rangle + 2 \kappa(N - \langle a^\dagger a \rangle)
\\[0.2cm]
\frac{\ud \langle a a \rangle}{\ud t} &=& 2 \mathcal{E} \langle a \rangle + 2 \kappa M_0 e^{-2i\Delta_{sp} t} - 2(\kappa + i \Delta_{cp}) \langle a a \rangle
\end{IEEEeqnarray*}
with $M_0=M_{t=0}$. 
We are interested in the steady-state. 
For the first two equations one readily finds
\begin{IEEEeqnarray}{rCl}
\langle a \rangle_\text{ss} &=& \frac{\mathcal{E}}{\kappa + i \Delta_{cp}} 
\\[0.2cm]
\langle a^\dagger a \rangle_\text{ss} &=& N + \frac{|\mathcal{E}|^2}{\kappa^2 + \Delta_{cp}^2}
\label{spump:ada}
\end{IEEEeqnarray}
However, the equation $\langle a a \rangle$  depends explicitly on time.
To proceed we set  $\langle a a \rangle = \langle a \rangle^2 + e^{-2 i \Delta_{sp} t} x(t)$ which will produce a time-independent equation for $x(t)$. Solving this equation then leads to 
\begin{equation}
\langle a a \rangle_\text{ss} = \frac{\kappa}{\kappa + i \Delta_{cs}} M_0 e^{2i \Delta_{sp}t} + \frac{\mathcal{E}^2}{[\kappa + i \Delta_{cp}]^2}
\end{equation}

We now return to the laboratory frame. 
This means we should multiply the result for $\langle a \rangle$ by $e^{-i \omega_p t}$ and the result for $\langle a a \rangle$ by $e^{-2i \omega_p t}$, leading finally to 
\begin{IEEEeqnarray}{rCl}
\label{eq:a_ss}
\langle a \rangle_\text{ss} &=& \frac{\mathcal{E}}{\kappa + i \Delta_{cp}}  e^{-i \omega_p t}	\\[0.2cm]
\label{eq:ada_ss}
\langle a^\dagger a \rangle_\text{ss} &=& N + \frac{|\mathcal{E}|^2}{\kappa^2 + \Delta_{cp}^2}			\\[0.2cm]
\label{eq:aa_ss}
\langle a a \rangle_\text{ss} &=& \frac{\kappa}{\kappa + i \Delta_{cs}} M_0 e^{-2i \omega_s t} + \frac{\mathcal{E}^2}{[\kappa + i \Delta_{cp}]^2} e^{-2i\omega_p t}
\end{IEEEeqnarray}

The Wigner function of the system will be Gaussian and thus is completely determined by the mean and variance of $a$ and $a^\dagger$. 
Define,
\begin{IEEEeqnarray}{rCl}
\bm{u} &=& (a,a^\dagger)		\\[0.2cm]
\bm{\alpha} &=& (\alpha,\alpha^*)\\[0.2cm]
\bm{\mu} &=& (\langle a \rangle, \langle a^\dagger \rangle) 		\\[0.2cm]
\Theta_{i,j} &=&  \frac{1}{2} \langle \{u_i,u_j\} \rangle - \langle u_i \rangle\langle u_j \rangle
\end{IEEEeqnarray}
Then, the most general Gaussian state of a single bosonic mode may be written as 
\begin{equation}\label{gen_Gaussian_W}
W = \frac{1}{\pi \sqrt{|\Theta|}} \exp\bigg\{ - \frac{1}{2} (\bm{\alpha} - \bm{\mu})^\dagger \Theta^{-1} (\bm{\alpha} - \bm{\mu}) \bigg\}
\end{equation}
The covariance matrix in the steady-state becomes
\begin{equation}\label{SS_Theta}
\Theta = \begin{pmatrix}
N+\nicefrac{1}{2} 	&	\tilde{M}_t		\\
\tilde{M}_t^* 		&	N + \nicefrac{1}{2}
\end{pmatrix}
\end{equation}
where 
\[
\tilde{M}_t = \frac{\kappa}{\kappa + i \Delta_{cs}} M_t
\]
This has the usual structure of a squeezing covariance matrix, but with a different $M$-parameter than the one expected from the bath. 
The difference is caused precisely by $\Delta_{cs}$, which is the mismatch between the cavity and the bath-induced frequencies.

%\subsection{Energy Flux}
%%
%The time derivative of the average energy is given by
%\begin{equation} \label{eq:AE}
%\frac{d\langle H\rangle}{dt} = \left\langle\frac{\partial H}{\partial t}\right\rangle + \tr[H\mathcal{D}_z(\rho)].
%\end{equation}
%In the steady state ($\frac{d\langle H\rangle}{dt}=0$), in which case we may derive the energy flux as
%\begin{equation} 
%\Phi_E=-\tr[H\mathcal{D}_z(\rho)]=\left\langle\frac{\partial H}{\partial t}\right\rangle.
%\end{equation}
%Now we can calculate the energy flux in the steady state as
%\begin{IEEEeqnarray}{rCl}
%\Phi_E &=& \omega_p [ \langle a^\dagger \rangle_{ss} \mathcal{E} e^{-i\omega_p t} + \langle a \rangle_{ss} \mathcal{E}^* e^{i\omega_p t} ] \\[0.2cm]
%\Phi_E &=& \frac{2 \kappa \omega_p |\mathcal{E}|^2}{\kappa^2+\Delta_{cp}^2}
%\end{IEEEeqnarray}

As discussed in the text, in the steady-state $\ud S/\ud t = 0$ so that $\Pi = \Phi$. 
At first it is not obvious that $\ud S/\ud t=0$, since the Wigner function $W$ depends explicitly on time, even in the steady-state. 
To verify this explicitly, one may note that for a general  Gaussian state of the form~(\ref{gen_Gaussian_W}),  Wigner entropy reads
\begin{equation}
S = \frac{1}{2} \ln|\Theta|  + 1+\ln\pi
\end{equation}
Even though the steady-state covariance matrix $\Theta$ in Eq.~(\ref{SS_Theta}) depends on time through $M_t$, its determinant depends only on $|M_t|^2$ and is hence time-independent. 
Consequently, we indeed have $\ud S/\ud t =0$.

\end{document}